\documentclass[%
 amsmath,amssymb,
 aps,
]{revtex4-1}

\usepackage{graphicx}
\usepackage{dcolumn}
\usepackage{bm}

\begin{document}

\preprint{APS/123-QED}

\title{Overset LES with Application to a Turbulent \\Channel Flow}

\author{R.A.D.~Akkermans}
\altaffiliation[]{Institute of Fluid Mechanics (ISM), TU Braunschweig, Germany.}

\author{R.~Ewert}%
\altaffiliation[]{Institute of Aerodynamics and Flow Technology, Dept. Technical Acoustics, DLR.}

\author{N.~Buchmann}
\altaffiliation[]{Institute of Fluid Mechanics (ISM), TU Braunschweig, Germany.}
\author{J.~Dierke}
\altaffiliation[]{Institute of Aerodynamics and Flow Technology, Dept. Technical Acoustics, DLR.}

\author{P.~Bernicke}
\altaffiliation[]{Institute of Fluid Mechanics (ISM), TU Braunschweig, Germany.}

%


\date{\today}

\begin{abstract}
In this contribution, we present an application of a novel perturbation approach (denoted as an Overset approach) to a generic turbulent channel flow. The derivation of the governing equations (the Non-Linear Perturbation Equations extended with fluctuating viscous terms) is presented as well as subgrid-scale modeling aspects which results in the Overset LES (OLES) method. The application of the Overset method is illustrated with a generic fully developed turbulent channel flow, the results of which show excellent agreement with reference Direct Numerical Simulation data. The ability of the overset method to correct for an imperfect background flow is furthermore shown.
\end{abstract}

\maketitle


\section{Introduction}
\begin{figure}[h!]
\centering
\includegraphics[width=0.8\textwidth]{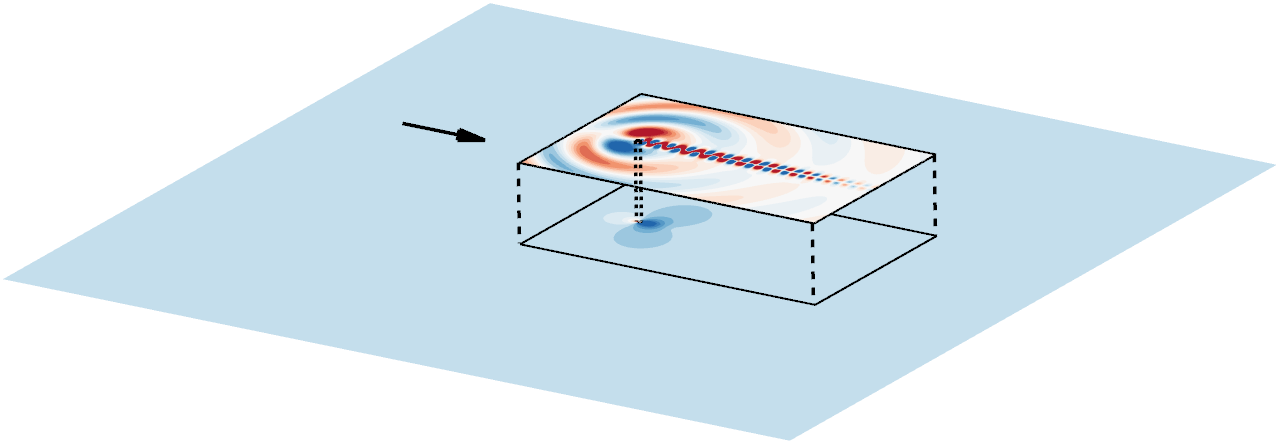}
 \caption{Schematic illustrating the overset approach (cylinder in uniform flow). The arrow indicates the undisturbed flow direction. The mean flow (lower plane) consists of a steady RANS background flow and the top layer corresponds to the perturbation analysis by overset DNS. Colors represent the mean pressure (lower plane) or pressure fluctuations (top plane).}
    \label{fig:overset2}
\end{figure}

Direct Numerical Simulations (DNS) provide the ability to simulate turbulent flows with the least amount of physical modeling. This is accompanied by a strong computational demand due to the scale disparity which restricts DNS often to simple geometries and moderate Reynolds numbers. Large-Eddy Simulation (LES) offer a way to reduce this computational demand by modeling the small non-resolved length scales (see, e.g., \cite{sagaut}).\\
\indent On the other side of the spectrum are approaches based on the Reynolds Averaged Navier-Stokes (RANS) equations. Such RANS based approaches are applicable for many engineering applications (i.e., complex geometries and high Reynolds numbers), however, involves significant modeling assumptions that fail to predict complex flow phenomena with high accuracy, e.g., pressure driven separation and the corresponding flow topology. Hybrid RANS/LES methods offer an attractive alternative to RANS or LES stand-alone methods where through a suitable combination of RANS and LES only LES is performed in the regions where it is needed and RANS in the remainder. For an overview concerning Hybrid LES/RANS approaches the reader is referred to \cite{froehlich}.\\
\indent In Computational Aeroacoustics (CAA) hybrid approaches are common in simulating acoustic propagation, where perturbation equations (such as the linearized Euler equations) are solved on top of a background flow. Viscous effects or nonlinearity are often neglected as they play a minor role for sound propagation over small distances and when considering small acoustic perturbations.\\
\indent In the current work, an extended version of PIANO (Perturbation Investigation of Aerodynamic NOise) is used. PIANO is a block-structured, high-order, low-dispersive and dissipative CAA code \cite{delfs}. PIANO solves perturbation equations, e.g., the Linearized Euler Equations (LEE), Acoustic Perturbation Equations (APE) \cite{ewertAPE}, or non-linear Euler Equations in perturbation form (PENNE) \cite{long} over a time-averaged background flow. The non-linear Euler equations have been augmented with fluctuating viscosity terms and therefore represent the fully non-linear Navier-Stokes equations formulated in perturbation form over a given base flow (see, e.g., \cite{ewert,ewert2014,mohsen}). In the majority of cases, the base flow constitutes a steady background flow obtained with a RANS simulation.\\
\indent In the above described approach, PIANO is applicable as a zonal-hybrid tool for scale-resolving simulation of the flow. In contrast to the so-called embedded hybrid approach, the current method is referred to as ``Overset'' to emphasize that a perturbation simulation is performed on top of a background flow. 
\begin{figure}[h!]
\centering
\includegraphics[width=0.95\textwidth]{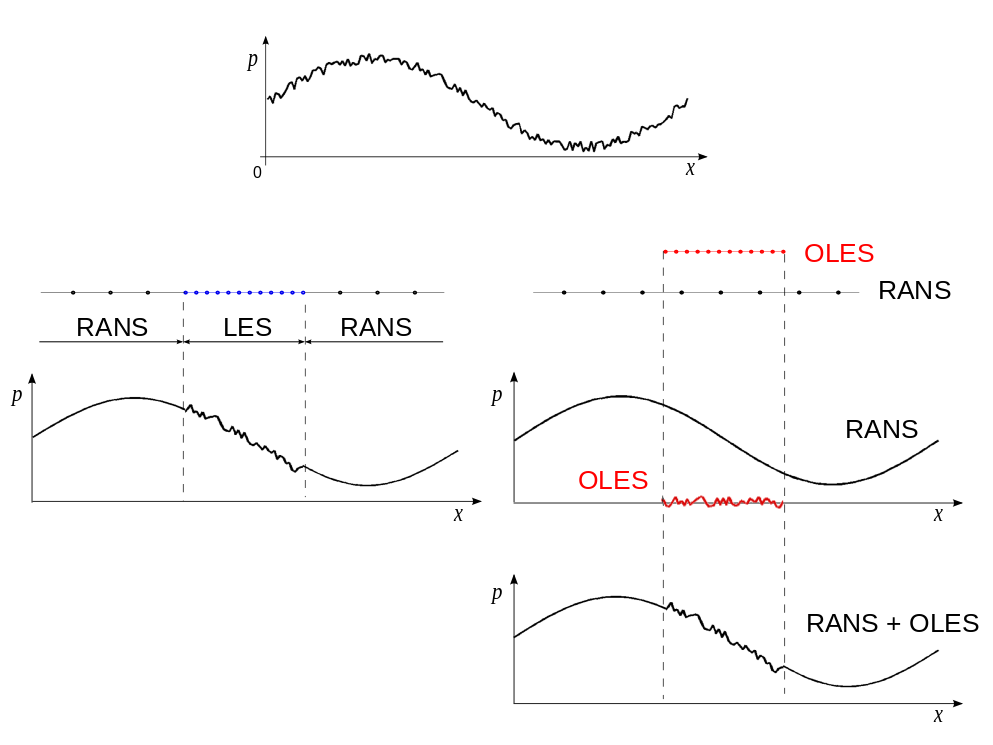}
\caption{Schematic illustration of (left) Embedded approach where for a single location one is either in a RANS or LES region and (right) Overset approach where there is a region where the RANS is overlapped with a (perturbation) Overset LES (OLES) region. Dots schematically represent the discretization of the solid line.}
    \label{fig:overset}
\end{figure}
In Fig.~\ref{fig:overset} the difference between the embedded (left) and overset approach (right) is illustrated. As opposed to Embedded LES, there exists an area where the RANS and LES overlap (red color in Fig.~\ref{fig:overset}) with an Overset LES (OLES) grid. As such, it constitutes a hybrid technique similar to the one proposed by Terracol \cite{terra}, however, formulated in primitive variables. An additional benefit of the here proposed method is that the computed nonlinear perturbations provide information about the correctness of the underlying base flow (i.e., a non-zero steady part of the perturbations should be interpreted as a correction of the underlying background flow). Possible applications are slat noise (i.e., wing with extended slat) or trailing-edge noise.\\
\indent The overset approach is further illustrated with the aid of a cylinder in uniform flow at low Reynolds number in Fig.~\ref{fig:overset2} (see, e.g., \cite{hrlm}). An application to a decaying Taylor-Green vortex flow is presented in Ref.~\cite{stab}. The cylinder is here indicated by the black circle. The steady background flow is depicted in the lower plane of Fig.~\ref{fig:overset2} and represents the mean pressure field around the cylinder. 
The fluctuating pressure obtained with the overset technique is presented in the top layer, showing strong hydrodynamic pressure fluctuations in the cylinder wake and weaker (as compared to the hydrodynamic) acoustic pressure fluctuations whose propagation direction is mainly normal to the uniform flow. This cylinder in uniform flow illustrates a DNC application of PIANO (i.e., simultaneously resolving the turbulent scales (DNS) including the radiated acoustic perturbations, see \cite{bailly}). As a comparison, PIANO in its traditional CAA-modus would use a steady background flow on top of which the acoustic propagation is performed, e.g., with the LEE.\\ 
\indent The present paper reports on the application of a novel perturbation approach (denoted as Overset approach) to a generic fully developed turbulent channel flow. The derivation of the Non-Linear Perturbation Equations (NLPE) with fluctuating viscous terms is presented as well as information related to the numerical method. The ability to simulate turbulent flow with the overset approach is illustrated with a turbulent channel flow. Results show an excellent agreement with reference DNS data of Moser et al. \cite{moser1}. The ability of the overset method to successfully correct for an imperfect background flow is furthermore shown.\\
\indent This paper is organized as follows. In Sect.~2 the derivation of the governing equations (the extended viscous NLPE) is presented in the second Section, followed by details concerning the numerical method in Section~3. Hereafter, in Sect.~4 the computational setup and results of the fully developed turbulent channel flow are presented. Finally, in Sect.~5 the main conclusions are summarized.\\

\section{Governing Equations}
\label{sec:govEq}
In the first subsection, the viscous non-linear disturbance equations are presented which are applicable in the so-called overset DNS approach. Hereafter, the extension is made to Overset LES (OLES) by presenting the subgrid-scale modeling is the second subsection. 
\subsection{Viscous Non-Linear Perturbation Equations}
The Navier-Stokes equations in primitive perturbation formulation form constitutes the basis of the here used governing equations. The Navier-Stokes equations for a compressible flow read
\begin{align}
&\frac{\partial\rho}{\partial t}+  v_i \frac{\partial\rho }{\partial x_i} +\rho \frac{\partial v_i }{\partial x_i}= 0  \label{eq:01a}&\\
&\frac{\partial v_i}{\partial t}+ v_j \frac{\partial v_i}{\partial x_j} +\frac{1}{\rho}\frac{\partial p}{\partial x_i} - \frac{1}{\rho}\frac{\partial \tau_{ij}}{\partial x_j}= 0  \label{eq:01b}&\\
&\frac{\partial p }{\partial t}+ v_i \frac{\partial p }{\partial x_i} +  \gamma p \frac{\partial v_i }{\partial x_i}  -(\gamma-1)\left[ \tau_{ij}\frac{\partial v_i }{\partial x_j} -\frac{\partial q_i }{\partial x_i}\right]= 0 \label{eq:01c}&
\end{align}
with density $\rho$, velocity vector $v_i$, and pressure $p$. The heat flux vector and viscous stress tensor are denoted by $q_i$ and $\tau_{ij}$, respectively. Furthermore, $\gamma$ represents the specific heat ratio. Einstein summation convention is implied for repeated indices. The use of primitive variables (i.e., density $\rho$, velocity $v_i$, and pressure $p$) has the advantage that it delivers directly the desired variable of interest in aeroacoustics: the fluctuating pressure (see, e.g., \cite{sesterhenn}). We are primarily interested in applying the Overset LES approach to subsonic problems, for which a formulation in primitive perturbation variables is suitable. Energy conservation of the mean-flow is guaranteed by a strongly conservative RANS simulation (see, e.g., \cite{dieter}). Preliminary test with a energy conserving skew-symmetric formulation of the NLPE do not indicate any improvement so far. However, to capture also supersonic simulation problems with possible shocks \cite{anderson}, also conservative perturbation formulations (NLDE) will be studied with PIANO in future work. Non-linear disturbance equations (NLDE) starting from  the Navier-Stokes equations in conservative form have been proposed by Morris et al. \cite{morris} and Liu \& Long \cite{long2}, or NLDE based LES by Terracol \cite{terra}.\\
\indent 
The perturbation form of the above Navier-Stokes equations are obtained by substitution of a decomposition of the variables into a background part and a fluctuating part, i.e., 
\begin{equation}\label{eq:02}
\rho=\rho^0+\rho^\prime, \;\; v_i=v_i^0+v_i^\prime, \;\; p=p^0+p^\prime, \;\; \tau_{ij}=\tau_{ij}^0+\tau_{ij}^\prime, \;\;\textrm{and}\;\;q_i=q^0+q^\prime. 
\end{equation}
The base flow variables are indicated by the superscript ``0'' and perturbations by a prime. Note that in Eqs.~(\ref{eq:02}), $\tau_{ij}^0$ and $\tau_{ij}^\prime$ denote viscous background flow and fluctuating viscous contributions, respectively, they do not represent turbulent Reynolds stresses. Upon substitution of the above decomposition into the compressible Navier-Stokes equations [Eqs.~(\ref{eq:01a}-\ref{eq:01c})], one obtains after some rearrangement
\begin{align}
&\frac{\partial \rho^\prime}{\partial t}+ \frac{\partial}{\partial x_i}( \rho^\prime v_i+ \rho^0 v_i^\prime) = m^0  \label{eq:03a}&\\
&\frac{\partial v_i^\prime}{\partial t}+ v_j \frac{\partial v^\prime_i}{\partial x_j} + v^\prime_j \frac{\partial v^0_i}{\partial x_j} -\frac{1}{\rho}\frac{\partial \sigma^\prime_{ij}}{\partial x_j} + \frac{\rho^\prime}{\rho\rho^0}\frac{\partial \sigma^0_{ij}}{\partial x_j}= f^0_i \label{eq:03b} &\\
&\frac{\partial p^\prime }{\partial t} + \frac{\partial}{\partial x_i}( p^\prime v_i+ p^0 v_i^\prime) -(\gamma-1)\left[ \sigma^\prime_{ij}\frac{\partial v_i }{\partial x_j} +\sigma^0_{ij}\frac{\partial v^\prime_i }{\partial x_j} -\frac{\partial q^\prime_i }{\partial x_i}\right]= \vartheta^0, \label{eq:03c}
\end{align}
where use has been made of the definition for the stress tensor $\sigma_{ij}:=\tau_{ij}-p\delta_{ij}$ (with $\sigma_{ij}=\sigma^0_{ij}+\sigma^\prime_{ij}$). In these equations, variables without superscript denote total variables and produce nonlinear terms when combined with primed variables. Terms containing only background flow contributions have been collected in the right-hand side residual terms and are given by
\begin{align}
&m^0 = -\left[\frac{\partial \rho^0}{\partial t}+ \frac{\partial \rho^0 v^0_j}{\partial x_i}\right] \label{eq:04a} &\\
&f^0_i = -\left[\frac{\partial v_i^0}{\partial t}+ v^0_j \frac{\partial v^0_i}{\partial x_j} +\frac{1}{\rho^0}\frac{\partial p^0}{\partial x_i} - \frac{1}{\rho^0}\frac{\partial \tau^0_{ij}}{\partial x_j}\right] \label{eq:04b}\\
&\vartheta^0 = -\left[\frac{\partial p^0 }{\partial t} + v_i^0\frac{\partial p^0}{\partial x_i} + \gamma p^0\frac{\partial v^0_i}{\partial x_i}  -(\gamma-1)\Big( \tau^0_{ij}\frac{\partial v^0_i }{\partial x_j}-\frac{\partial q^0_i }{\partial x_i}\Big)\right].\label{eq:04c}
\end{align}
This set of equations [Eqs.~(\ref{eq:03a}-\ref{eq:04c})], the viscous Non-Linear Perturbation Equations (NLPE), are applicable to simulate laminar and turbulent sound generation in the manner of a Direct Numerical Simulation (DNS). The base flow part of the viscous stress tensor and heat flux vector are denoted by $\tau^0_{ij}$ and $q_i^0$, respectively. Therefore the right-hand side residuals represent residual turbulent viscous stress and heat flux whose exact definition is related to the base flow definition. For a base flow obtained from a steady compressible RANS simulation, where the variables indicated by ``0'' are Favre averaged, one can write 
\begin{align}
&\frac{\partial \rho^0}{\partial t}+ \frac{\partial \rho^0 v^0_j}{\partial x_i}  = 0  \label{eq:05a} &\\
&\frac{\partial v_i^0}{\partial t}+ v^0_j \frac{\partial v^0_i}{\partial x_j} +\frac{1}{\rho^0}\frac{\partial p^0}{\partial x_i} - \frac{1}{\rho^0}\frac{\partial \tau^0_{ij}}{\partial x_j}= \frac{1}{\rho^0}\frac{\partial (\tau^0_{ij})_\textrm{turb}}{\partial x_j} \label{eq:05b} &\\
&\frac{\partial p^0 }{\partial t} + v_i^0\frac{\partial p^0}{\partial x_i} + \gamma p^0\frac{\partial v^0_i}{\partial x_i}  -(\gamma-1)\Big( \tau^0_{ij}\frac{\partial v^0_i }{\partial x_j}-\frac{\partial q^0_i }{\partial x_i}\Big)=\nonumber \\
&\qquad\qquad\qquad\qquad\qquad\qquad (\gamma-1)\Big[ (\tau^0_{ij})_\textrm{trub}\frac{\partial v^0_i }{\partial x_j}-\frac{\partial (q^0_i)_\textrm{trub} }{\partial x_i}\Big].\label{eq:05c}
\end{align}
Finally, the residual right-hand side terms become for a steady compressible RANS base flow
\begin{align}
&m^0  = 0 \label{eq:06a} &\\
&f_i^0=- \frac{1}{\rho^0}\frac{\partial (\tau^0_{ij})_\textrm{turb}}{\partial x_j} \label{eq:06b} &\\
&\vartheta^0=(\gamma-1)\Big[ (\tau^0_{ij})_\textrm{trub}\frac{\partial v^0_i }{\partial x_j}-\frac{\partial (q^0_i)_\textrm{trub} }{\partial x_i}\Big]. \label{eq:06c}
\end{align}
The residual right-hand source terms, represent the residual turbulent viscous stresses and heat fluxes, whose exact definition depends on the definition of the base flow.\\
\indent For the case that the base flow is obtained from a steady RANS simulation, the storage requirement for background flow components can be reduced in the following way. With time derivatives equal zero for a steady RANS, the momentum Eq.~(\ref{eq:05b}) can be used to rewrite the divergence of the background stress tensor $\sigma^0_{ij}$, i.e. 
\begin{equation}\label{eq:07}
\frac{1}{\rho^0}\frac{\partial \sigma^0_{ij}}{\partial x_j}=   v^0_j \frac{\partial v^0_i}{\partial x_j} -\frac{1}{\rho^0}\frac{\partial (\tau^0_{ij})_\textrm{turb}}{\partial x_j}.  
\end{equation}
The number of additional background flow variables that need to be permanently stored reduces from 13 (6 for the mean-flow stress tensor, three for the divergence of the mean-flow stress tensor, and 4 for the residual right-hand side terms) to 10 (6 for the mean-flow stress tensor, three for the divergence of the mean-flow stress tensor, and 1 for $\vartheta^0$). The extended viscous NLPE (with memory-reduced momentum equations) finally becomes 
\begin{align}
&\frac{\partial \rho^\prime}{\partial t}+ \frac{\partial}{\partial x_i}( \rho^\prime v_i+ \rho^0 v_i^\prime) = 0 \label{eq:08a} &\\
&\frac{\partial v_i^\prime}{\partial t}+ v_j \frac{\partial v^\prime_i}{\partial x_j} + v^\prime_j \frac{\partial v^0_i}{\partial x_j}+ \frac{\rho^\prime}{\rho} v^0_j \frac{\partial v^0_i}{\partial x_j} - \frac{1}{\rho}\frac{\partial \sigma^\prime_{ij}}{\partial x_j} = \hat{f}^0_i \label{eq:08b} &\\
&\frac{\partial p^\prime }{\partial t} + \frac{\partial}{\partial x_i}( p^\prime v_i+ p^0 v_i^\prime) -(\gamma-1)\left[ \sigma^\prime_{ij}\frac{\partial v_i }{\partial x_j} +\sigma^0_{ij}\frac{\partial v^\prime_i }{\partial x_j} -\frac{\partial q^\prime_i }{\partial x_j}\right]= \vartheta^0.\label{eq:08c}
\end{align}
where
\begin{equation}\label{eq:09}
\hat{f}^0_i=-\frac{1}{\rho^0}\frac{\partial (\tau^0_{ij})_\textrm{turb}}{\partial x_j}+ \frac{\rho^\prime}{\rho^0\rho}\frac{\partial (\tau^0_{ij})_\textrm{turb}}{\partial x_j}=-\frac{1}{\rho}\frac{\partial (\tau^0_{ij})_\textrm{turb}}{\partial x_j}.
\end{equation}
The above presented extended viscous NLDE are preferred over the original viscous NLDE as they require less memory storage (see \cite{ewert}). Equations~(\ref{eq:08a}-\ref{eq:09}) reduce to the PENNE equations as proposed by Long \cite{long} when the viscous terms are neglected.\\
\indent The computed perturbations provide information concerning the correctness of this underlying background flow, i.e., a non-zero steady part of the perturbation provides a correction for the RANS background flow. Note that for aeroacoustic applications, the correct temporal evolution of the pressure fluctuations is of paramount importance and the (residual) mean pressure should be subtracted as this masks the acoustic content
\begin{equation}
p^\prime=p_{\textrm{PIANO}}^\prime-\overline{p_{\textrm{PIANO}}^\prime}, 
\end{equation}
where $p_{\textrm{PIANO}}^\prime$ is the computed perturbation pressure from PIANO.
\subsection{Overset Large-Eddy Simulations: Subgrid-Scale Models}
The in the previous subsection described Navier-Stokes equations can be made usable as Overset LES (OLES) by extension with suitable subgrid-scale (SGS) models for the stress and heat flux contributions [i.e., $(\tau_{ij})_\textrm{sgs}$ and $\vartheta_\textrm{sgs}$, respectively]. Note that then the equation variables implicitly represent Favre-average variables (i.e., density-weighted averaging). For the flow variables this yields  
\begin{equation}\label{eq:09aa}
\widetilde{\rho}=\overline{\rho}, \;\; \widetilde{v}_i=\overline{\rho v_i}/\overline{\rho}, \;\;\textrm{and}\;\; \widetilde{p}=\overline{p},
\end{equation}
where spatial filtering is denoted by the overbar and Favre averaging by a tilde. The subgrid-scale terms are defined in terms of Favre-averaged flow variables, for the SGS stress term this yields
\begin{equation}\label{eq:09ab}
(\tau_{ij})_\textrm{sgs} = \widetilde{\rho}\Big( \widetilde{v_iv_j}-\widetilde{v}_i\widetilde{v}_j\Big), 
\end{equation}
where the exact definition depends upon the used SGS model. Two SGS models will be described, explicit spatial filtering and the classical Smagorinsky (see, e.g., \cite{sagaut}). Furthermore, an SGS-model based on the so-called Flow Simulation Methodology (FSM, see \cite{froehlich}) is discussed.
\subsubsection{Explicit Filtering}
Spatial filtering plays a important role in CAA codes. Normally CAA codes have low dissipative finite difference schemes, and besides this (artificial) numerical dissipation there is often no physical dissipation in the governing equations when using, e.g., the LEE. Filtering is utilized to remove grid to grid oscillations, i.e., spurious high-wavenumber oscillations. Furthermore, it can also act as a SGS model by removing the high-frequency small scales through explicit spatial filtering of flow variables. Such an approach is motivated by considering the smallest scales mainly responsible for dissipative action which can be modeled by removing the high frequency part of the solution. Implicit LES approaches based on explicit filtering has been employed by Visbal et al. \cite{visbal1}, Rizzetta et al. \cite{visbal2}, and Marsden et al. \cite{marsden}. An implicit LES approach based on spatial filtering (as practiced e.g., \cite{marsden}), is easily implemented by utilizing the finite-difference filters that are available in the PIANO code. 
\subsubsection{Smagorinsky Model}
The deviatoric part of the stress tensor is modeled with the eddy viscosity hypothesis, i.e.,
\begin{equation}\label{eq:10}
\tau_{ij}-\frac{1}{3}\delta_{ij}\tau_{kk}= -2\nu_\textrm{sgs} \widetilde{S}_{ij}, 
\end{equation}
where $\widetilde{S}_{ij}$ denotes the resolved-velocity strain-rate tensor and $\nu_\textrm{sgs}$ the eddy viscosity. The latter is computed with the classical Smagorinsky model which reads \begin{equation}\label{eq:11}
\nu_\textrm{sgs} = (C_S\Delta)^2 |\widetilde{S}_{ij}|. 
\end{equation}
Here, $C_S$ represents the Smagorinsky constant and $\Delta$ the filter width which is related to the grid size, here taken to be $\Delta=(\Delta_x\Delta_y\Delta_z)^{1/3}$. Combination of the last two equations reveals that the modeled stress tensor is proportional with the resolved strain-rate tensor squared. For the perturbation approach, an additional term appears on the right hand side of the momentum equation  $\frac{1}{\rho}\frac{\partial \tau^\textrm{sgs}_{ij}}{\partial x_j}$ with $\tau^\textrm{sgs}_{ij}$
\begin{equation}\label{eq:10a}
\tau^\textrm{sgs}_{ij}= -2\nu_\textrm{sgs} \widetilde{S}_{ij}=-2(\nu^0_\textrm{sgs}+\nu^\prime_\textrm{sgs}) (\widetilde{S}^0_{ij} + \widetilde{S}^\prime_{ij}),
\end{equation}
where use has been made from the relations $\nu_\textrm{sgs}=\nu^0_\textrm{sgs}+\nu^\prime_\textrm{sgs}$ and $\widetilde{S}_{ij}=\widetilde{S}^0_{ij} + \widetilde{S}^\prime_{ij}$, i.e., a similar decomposition as Eqs.~(\ref{eq:02}). Some rearrangement yields
\begin{equation}\label{eq:10b}
\tau^\textrm{sgs}_{ij}=-2\nu^0_\textrm{sgs}\widetilde{S}^0_{ij}-2\nu^\prime_\textrm{sgs}\widetilde{S}^0_{ij}-2\nu_\textrm{sgs}\widetilde{S}^\prime_{ij}, 
\end{equation}
where variables without a superscript denote total variables. The first term on the right-hand side is a background flow contribution to the SGS and therefore added to Eq.~(\ref{eq:09}), whereas the remaining terms represent linear and non-linear SGS contributions and are shifted to the left-hand side of momentum equation Eq.~(\ref{eq:08b}).\\   
\indent The subgrid-scale heat flux term $\vartheta^0_\textrm{sgs}$ in the energy equation is modeled in a similar way as is done above, by a proportionality coefficient $\kappa_\textrm{sgs}$ which is related to the eddy viscosity by
\begin{equation}\label{eq:12}
\kappa_\textrm{sgs} = \frac{c_p \mu_\textrm{sgs}}{\textrm{Pr}_\textrm{sgs}}, 
\end{equation}
with $\textrm{Pr}_\textrm{sgs}$ the subgrid-scale Prandtl number (taken equal to 0.6 \cite{adams}). The turbulent eddy viscosity is obtained as described above. The subgrid-scale heat flux term $\vartheta_\textrm{sgs}$ is modeled as
\begin{equation}\label{eq:13}
\vartheta_\textrm{sgs}= \kappa_\textrm{sgs} \frac{\partial T^\prime}{\partial x_j}+\frac{\mu-\mu^0}{\mu^0}q^0_j, 
\end{equation}
and added to the right hand side source term $\vartheta^0$ [Eq.(\ref{eq:06c})].

\subsubsection{Flow Simulation Methodology}
A further SGS model is presented that is based on the Flow Simulation Methodology (FSM, originally developed by Speciale \cite{speciale}) , i.e.  
\begin{equation}\label{eq:11b}
\tau^\textrm{mod}_{ij}=f_\Delta\Big(\frac{\Delta}{\ell_K}\Big)\tau^\textrm{RANS}_{ij},
\end{equation}
with $\Delta$ proportional to grid size and $\ell_K$ represents the Kolmogorov length scale. The ratio of these two variables is used to estimate the ``distance to DNS'' \cite{froehlich} with $f_\Delta$ the contribution function that is used to damp the contributions from the RANS model. For a course grid ($\Delta/\ell_K\gg1$) the RANS model is fully used and the contribution function equals one, for a DNS-like grid (i.e., $\Delta/\ell_K \approx 1$) the contribution function goes to zero.\\
\indent 
The FSM adaptation reads
\begin{equation}
\mu^\textrm{mod}_{ij}=f_\Delta\Big(\frac{\Delta}{\ell_K}\Big)\mu^\textrm{RANS}_{ij},
\end{equation}
with $\mu_{ij}=\rho \nu_{ij}$. The RANS turbulent viscosity $\nu^\textrm{RANS}_{ij}$ is estimated with the variables turbulent kinetic energy $k$ and dissipation $\varepsilon$. The modeled viscosity $\mu^\textrm{mod}_{ij}$ then becomes
\begin{equation}
\mu_{ij}^\textrm{mod}=F\Big(\Delta,\nu,k, \varepsilon\Big)=\mu_{ij}^{mod,0}+\mu_{ij}^{mod,\prime}.
\end{equation}
Therefore, the fluctuating part of the modeled viscosity equals zero, i.e., $\mu_{ij}^{mod,\prime}=0$. The turbulent viscous stress becomes
\begin{equation}
\tau^{t}_{ij}=\mu_{ij}^\textrm{mod}(\widetilde{S}^{0}_{ij}+\widetilde{S}^{\prime}_{ij}),
\end{equation}
so that for the fluctuating stress tensor it follows
\begin{equation}
\tau^{\prime}_{ij}=\tau^{t}_{ij}-\mu^\textrm{mod}\widetilde{S}^{0}_{ij}=\mu^\textrm{mod}\widetilde{S}^{\prime}_{ij}.
\end{equation}
As opposed to the Smagorinsky SGS-model, the FSM based SGS-model is linear proportional to the resolved strain-rate tensor.

\section{Numerical Method}
The in Section~2 presented governing equations are solved for the results in the remainder of this paper. The CAA-code PIANO \cite{delfs} is a structured code, based on curvilinear, multi-block grids. In addition to the in Section~2 presented viscous NLPE equations, it also supports the computation of sound propagation by the Linearized Euler Equations (LEE), Acoustic Perturbation Equations (APE) or non-linear Euler equations in primitive disturbance form (PENNE) as governing equations. As is fairly standard in aeroacoustics, optimized finite-difference schemes are used for the numerical discretization. Spatial gradients are approximated with the Dispersion Relation Preserving (DRP) scheme, as proposed by Tam and Webb \cite{tam}. Explicit time advancement is achieved by the 4th-order low-dispersion Runge-Kutta (LDDRK) algorithm, as proposed by Hu et al. \cite{hu}. Artificial selective damping  as well as different filters provides the possibility to remove contaminated high-frequency wave components.\\  
\indent New terms in the equations, as compared to PIANO in its traditional CAA modus, are the viscous and heat flux terms. These additional terms involve second derivative operators which could be tackled by the successive application of the first derivative operator. The computation of viscous terms mainly involves computation of divergence of vectors and are calculated as such. The routine based on divergence calculations also serves as a building block for the anticipated future implementation of a full conservative formulation of the governing equations. The interested reader is referred to Moghadam~\cite{mohsen} for a detailed description of the developed divergence computation.\\  

\section{Overset LES Results of a Turbulent Channel Flow}
The fully developed turbulent channel flow at $\textrm{Re}_\tau=395$ serves as a test case for the overset LES method, where $\textrm{Re}_\tau$ is the friction based Reynolds number based on the mean wall-friction velocity $u_\tau$. A comparison will be made mainly with corresponding DNS reference data from Moser et al. \cite{moser1}.
\subsection{Computational Setup}
The simulations have been carried out in a computational box of $L_x \times L_y \times L_z$ equal to $4\delta \times 2\delta \times 2\delta$. Here, the $x$- and $z$-axes are in the streamwise and spanwise direction, respectively. The $y$-axis is the wall-normal direction, perpendicular to the plane spanned by the $x,z$-axes. The number of grid points $N_x \times N_y  \times N_z$ equals $165 \times 193  \times 192$. Periodic boundary condition were enforced on the streamwise and spanwise directions. The channel walls were taken to be no-slip, isothermal and normal pressure gradient boundary conditions. The pressure gradient in the streamwise direction enters the simulation through the background flow (cf. the residual right-hand side term for the momentum equations). The computational mesh is equally spaced in the $x$- and $z$-direction ($\Delta x^+=\Delta z^+\approx 10$). A stretching has been applied to the wall-normal direction so as to have the minimum spacing at the wall of $\Delta y_{min}^+\approx 1$ and a maximum at the channel centerline of $\Delta y_{max}^+\approx 12$. The simulations were performed on the North-German Supercomputing Alliance (HLRN) cluster and were parallelized on 24 processor for the used computational domain. The timestep was set equal to $\Delta t=1 \cdot10^{-3}$ where typically about 600.000 timesteps were computed.\\
\indent The simulations have been initiated with the aforementioned RANS background flow, obtained with the DLR TAU-code (see, e.g., \cite{dieter}). It constitutes a 3D RANS simulation with the Reynolds stress model JHh-v2 (see, e.g., \cite{cecora}). With the FRPM code (see, e.g., \cite{ewertFRPM}) realistic turbulent fluctuations (that correspond to the turbulent kinetic energy distribution provided by the background flow) were generate and added to the RANS background flow to accelerate the development of turbulence. Typically a total of 45 convective time units were simulated with the Overset approach, and after an initial transient, the last 25 convective time units were sampled for statistics.\\
\subsection{RANS Background Flow}
\begin{figure}[h]
  \centering
\includegraphics[width=0.6\textwidth]{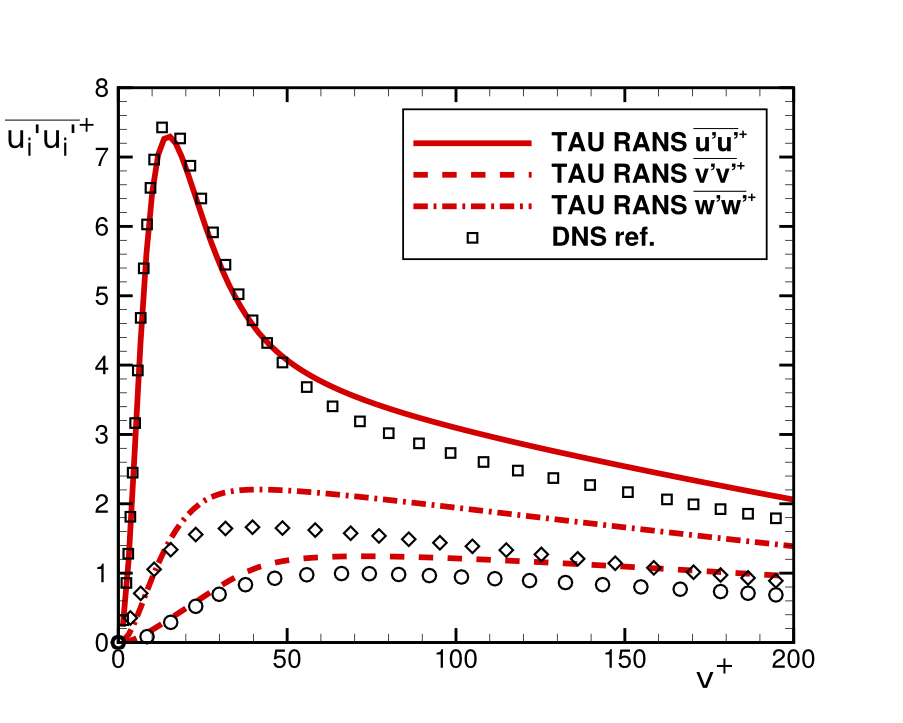}
  \caption{Non-dimensional Reynolds stress components of the steady RANS background flow together with DNS reference data (Moser et al. \cite{moser1}). From top to bottom $i=1$ ($\overline{u^\prime u^\prime}^+$), $i=3$ ($\overline{w^\prime w^\prime}^+$), and $i=2$ ($\overline{v^\prime v^\prime}^+$), respectively.}
  \label{fig:figRans2}
\end{figure}

First, the utilised steady background flow is presented obtained with the DLR TAU-code.   
\begin{figure}[ht!]
  \begin{minipage}[b]{0.5\textwidth}
   \includegraphics[width=\textwidth]{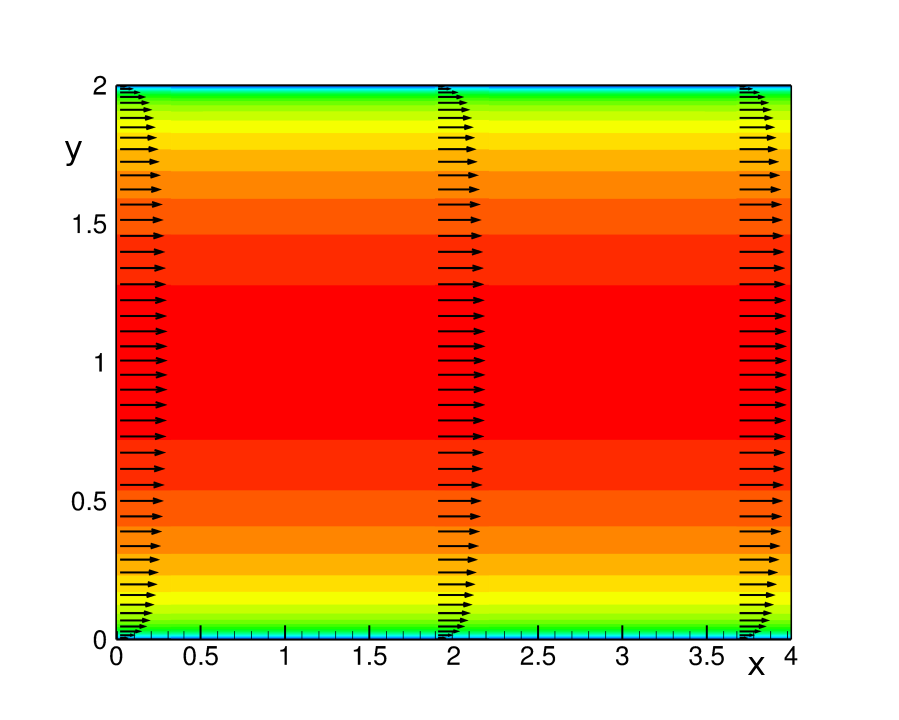}
    \end{minipage}
  \hfill
  \begin{minipage}[b]{0.5\textwidth}    
  \includegraphics[width=\textwidth]{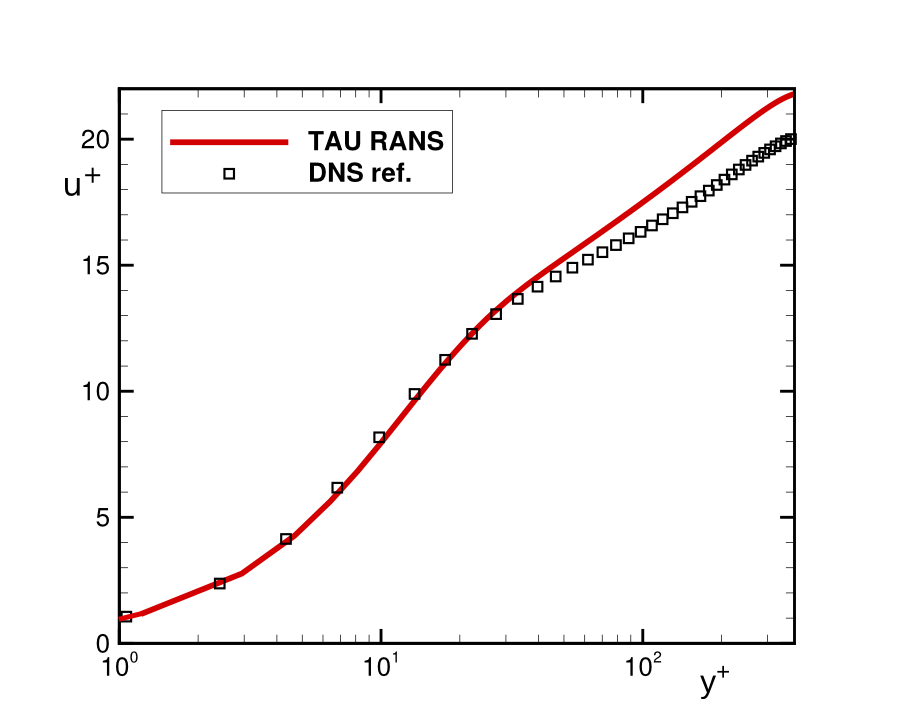}
  \end{minipage} 
  \caption{(left) $xy$-plane slice of steady RANS background flow showing contour and vectors of the streamwise velocity component. (Right) Non-dimensional mean velocity profile $u^+$ of the steady RANS background flow together with DNS reference data (Moser et al. \cite{moser1}).}
  \label{fig:figRans}
\end{figure}
Figure~\ref{fig:figRans}~(left) shows contours and vectors of the streamwise velocity component, where the typically full velocity profile of a turbulent channel flow can be appreciated. In the right plot of Fig.~\ref{fig:figRans}, the non-dimensional mean velocity $u^+$ profile is depicted for the RANS background flow together with the one for the DNS reference data. Clearly, the the logarithmic region of the non-dimensional mean velocity profile $u^+$ is substantially over-predicted.\\
\indent In Fig.~\ref{fig:figRans2} the normal Reynolds stress components given by the Reynolds stress model JHh-v2 (see, e.g., \cite{cecora}) are depicted as function of wall units. 
Comparison of the reference DNS data (Moser et al. \cite{moser1}) with the steady TAU RANS reveals a qualitative agreement with significant quantitative differences. In section~2, it was claimed that the overset approach is able to correct for such deficits in the background flow [see Figs.~\ref{fig:figRans}~(right) and \ref{fig:figRans2}], provided sufficient resolution of the overset approach. This would show up as a non-zero mean part of the calculated fluctuations (of the overset LES).\\ 

\subsection{Validation against DNS}
\begin{figure}[h]
  \centering
\includegraphics[width=0.65\textwidth]{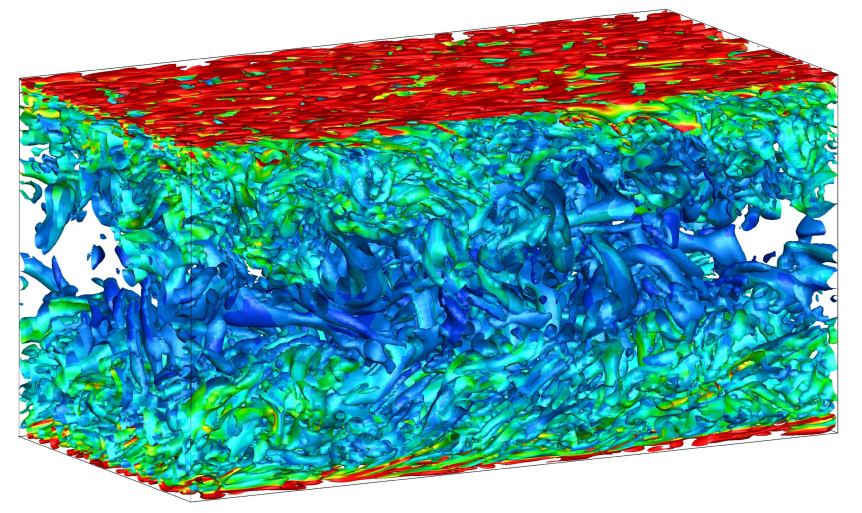}
  \caption{Snapshot of the $Q$-criterion for the fully developed turbulent channel flow at $\textrm{Re}_\tau=395$.}
  \label{fig:vortexshed}
\end{figure}

\begin{figure}[h]
  \begin{minipage}[b]{0.5\textwidth}
  \includegraphics[width=\textwidth]{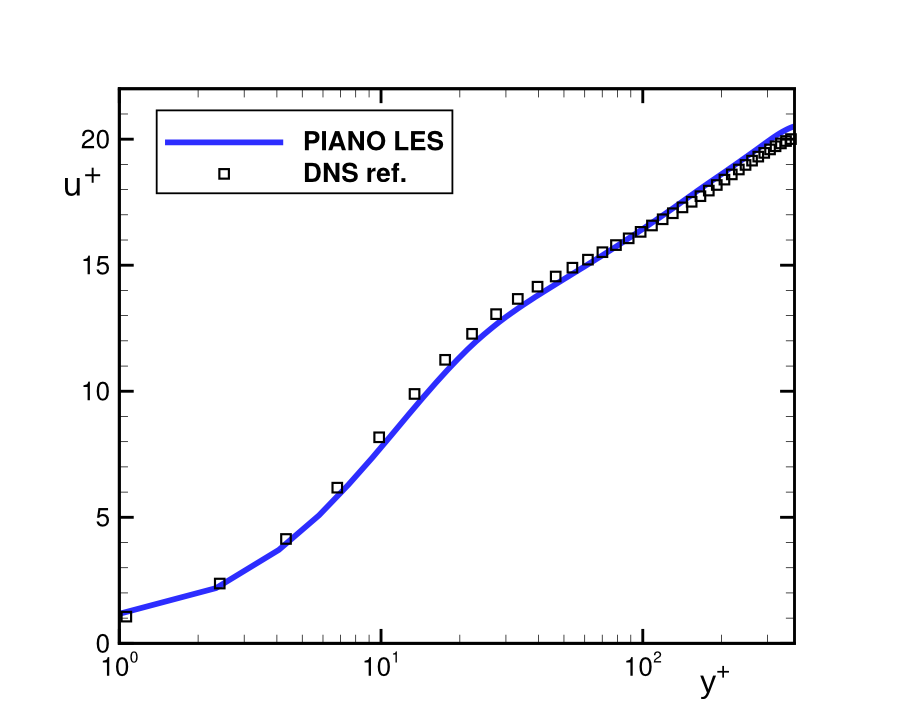}
    \end{minipage}
  \hfill
  \begin{minipage}[b]{0.5\textwidth}    
  \includegraphics[width=\textwidth]{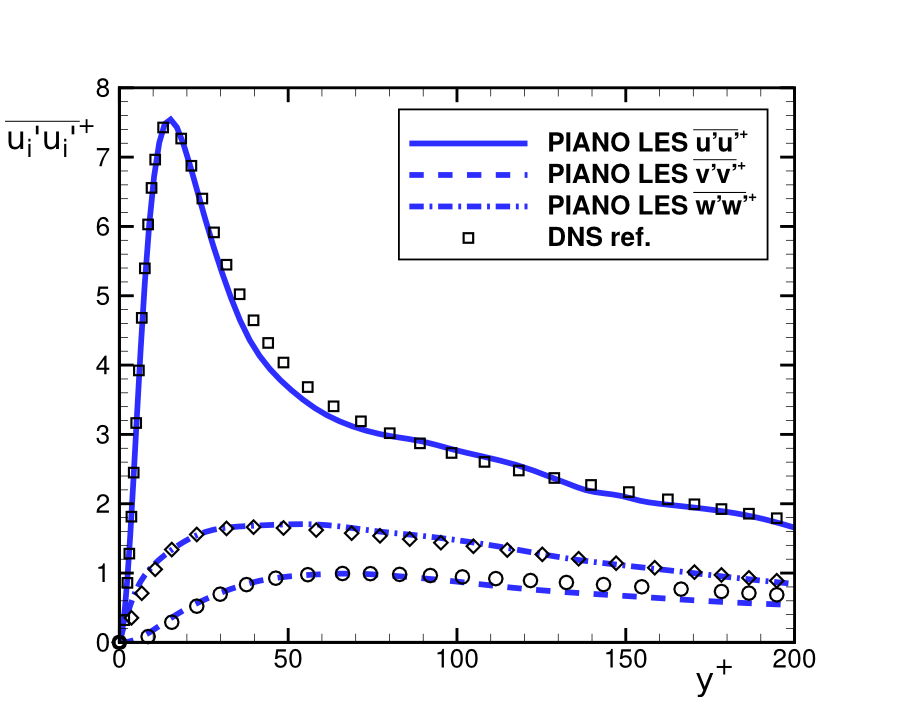}
  \end{minipage} 
  \caption{Fully developed turbulent channel flow at $\textrm{Re}_\tau=395$. (left) Non-diemnsional mean velocity profile $u^+$ and (right) non-dimensional Reynolds stress components for from top to bottom $i=1$ ($\overline{u^\prime u^\prime}^+$), $i=3$ ($\overline{w^\prime w^\prime}^+$), and $i=2$ ($\overline{v^\prime v^\prime}^+$). DNS reference data from Moser et al. \cite{moser1}.}
  \label{fig:figLES}
\end{figure}

In Fig.~\ref{fig:vortexshed} a snapshot of vortical structures (visualized with the $Q$-criterion) is depicted for the fully developed turbulent channel flow. 
Near the walls, small vortical structure are seen with increasing vortex size towards the centerline of the channel.\\
\indent The universal velocity profile $u^+$  of the fully developed channel flow obtained with the Overset LES method is displayed in Fig.~\ref{fig:figLES}~(left). It compares the PIANO simulation with a reference DNS of Moser et al. \cite{moser1}, illustrating that the logarithmic region is correctly captured. In the right plot of Fig.~\ref{fig:figLES} the normal Reynolds stresses are depicted for this turbulent channel flow. Comparison between the PIANO and DNS reference data reveals an excellent quantitative agreement. Peak value and position of, e.g., the streamwise Reynolds stress component are captured very well. 
Comparison of the RANS computed mean velocity and Reynolds stresses [Fig.~\ref{fig:figRans}~(right) and Fig.~\ref{fig:figRans2}, respectively] and the PIANO results for the mean velocity and Reynolds stresses (see Fig.~\ref{fig:figLES}) reveals that PIANO indeed is able to correct an imperfect background flow (provided sufficient resolution).

\section{Conclusions}
Hybrid RANS/LES methods offer an attractive alternative to stand-alone RANS or LES methods where LES is performed in the regions where it is needed and RANS in the remainder. A reduction in computational demand results as compared to stand-alone LES, facilitating the computation of more complex geometries and higher Reynolds numbers. In Computational Aeroacoustics such hybrid approaches are common, where the acoustic propagation is computed with a perturbation approach (such as linearized Euler equations) on top of a background flow.\\ 
\indent In this contribution, we have presented an application of a novel perturbation approach (denoted as an Overset approach) to a generic turbulent channel flow. The derivation of the Non-Linear Perturbation Equations extended with fluctuating viscous terms was presented as well as subgrid-scale modeling aspects which results in the Overset LES (OLES) method. The application of the Overset method was illustrated with a generic fully developed turbulent channel flow, the results of which show excellent agreement with reference Direct Numerical Simulation validation data from Moser et al. \cite{moser1}. Furthermore, the ability of the overset method to successfully correct for an imperfect background flow was shown.\\

\section*{Acknowledgement}
This work was funded by the Deutsche Forschungsgemeinschaft DFG (German Research Funding Organisation) in the framework of the Collaborative Research Centre CRC880 (SFB880). Computational resources were provided by the North-German Supercomputing Alliance HLRN, which is gratefully acknowledged.

\end{document}